\documentclass[10pt, conference]{IEEEtran}

\usepackage{url}

\ifCLASSINFOpdf
\else
\fi

\usepackage{graphicx}
\usepackage{color}   % just for \remark
\usepackage{subfigure}
\usepackage{pdfpages}
\usepackage{blindtext}
\usepackage{multirow}

\newcommand{\mahmoud}[1]{\textcolor{blue}{\em #1}}

\newcommand{\ignore}[1]{}

\begin{document}
%\title{Poster: Validation of Internal Meters of Mobile Android Devices}
\title{\vspace{-3mm}Validation of Internal Meters of Mobile Android Devices}


\author{\IEEEauthorblockN{Mahmoud A. Bokhari$^{1,2}$, Yuanzhong Xia$^1$, Bo Zhou$^1$, Brad Alexander$^1$, Markus Wagner$^1$}
%\author{\IEEEauthorblockN{(author information removed)}
\IEEEauthorblockA{
	\begin{tabular}[t]{@{}c@{}}
$^1$Optimization and Logistics Group\\
Computer Science School\\
University of Adelaide, Australia\\
\\
	\end{tabular}\nobreak\qquad
	\begin{tabular}[t]{@{}c@{}}
$^2$Computer Science Department\\
Taibah University\\
Medina, Kingdom of Saudi Arabia
	\end{tabular}\vspace{-4mm}
	\\$\{$mahmoud.bokhari,bradley.alexander,markus.wagner$\}$@adelaide.edu.au, $\{$yuanzhong.xia,bo.zhou$\}$@student.adelaide.edu.au
}
\ignore{\IEEEauthorblockA{
	\begin{tabular}[t]{@{}c@{}}
		\affaddr{$^1$Optimization and Logistics group}\\
		\affaddr{Computer Science School}\\
		\affaddr{University of Adelaide, Australia}\\
		\affaddr{}\\
	\end{tabular}\nobreak\qquad
	\begin{tabular}[t]{@{}c@{}}
		\affaddr{$^2$Computer Science Department}\\
		\affaddr{Taibah University}\\
		\affaddr{Medina, Kingdom of Saudi Arabia}
	\end{tabular}\vspace{-4mm}
	\\mahmoud.bokhari@adelaide.edu.au, yuanzhong.xia@student.adelaide.edu.au, bo.zhou@student.adelaide.edu.au, bradley.alexander@adelaide.edu.au, markus.wagner@adelaide.edu.au
}
}
}

\ignore{
\and
\IEEEauthorblockN{Authors Name/s per 2nd Affiliation (Author)}
\IEEEauthorblockA{line 1 (of Affiliation): dept. name of organization\\
line 2: name of organization, acronyms acceptable\\
line 3: City, Country\\
line 4: Email: name@xyz.com}
}


%Telephone: (800) 555--1212, Fax: (888) 555--1212}


\maketitle

\renewcommand\IEEEkeywordsname{Keywords}
\begin{abstract}
In this paper we outline our results for validating the precision of the internal power meters of smart-phones under different workloads. We compare its results with an external power meter. This is the first step towards creating customized energy models on the fly and towards optimizing battery efficiency using genetic program improvements. Our experimental results indicate that the internal meters are sufficiently precise when large enough time windows are considered. 

\end{abstract}


% \fi
\IEEEpeerreviewmaketitle

%\vspace{-1mm}
\section{Introduction}%\vspace{-1mm}
The energy requirements of modern smart-phones continue to increase as a result of various factors. Among these factors are the growing demands on their performance, users' needs such as smooth experience and longer operation times. However, the available energy provided by a smart-phone is restricted by its battery size. Optimizing energy consumption requires accurate measurements and estimation of the smart-phone energy usage patterns, as well as solutions which balance multiple (potentially conflicting) requirements. Therefore, improving the battery life starts from measuring the energy consumption of the phone.

We aim to implement an energy profiler that generates event-based energy consumption models on the fly using the smart-phones' battery fuel gauges. The use of built-in meters enables the use of online-learned and adaptive models that can be applied under varying conditions, accounting for each user's usage patterns~\cite{falaki:diversityUsagePattern,Shye:utilization1}, phone configuration, and battery age. Furthermore, it facilitates battery optimization processes. %Currently, we measure energy consumption using internal and external power meters while benchmarks directly on the phone. 
Although external power meters are golden standards \cite{energyConsumptionBook}, they are costly, restricted to laboratory conditions and hard to setup.  
However, before we can solely rely on internal meters, it crucial to validate their precision.

In this work we report on the results of a series of experiments on two different Android devices and compare the readings of their internal energy meters with an external meter. Our experiments show that using these internal meters is sufficient for detecting differences in energy consumption.

\section{Methodology}
In this section, we outline our software and hardware test harnesses. 

\subsection{Hardware}
Modern mobile phones are equipped with battery fuel gauge chips that report the voltage, current and remaining energy within the battery \cite{android:androidManual}.
The target devices for our experiments are HTC Nexus 9 and Motorola Nexus 6. They are equipped with the Maxim MAX17050 fuel gauge chip that compensates measurements for temperature, battery age and load \cite{maxim}. To validate the internal power meter measurements, we insert a Yoctopuce Yocto-Watt~\cite{yoctowatt} external meter between the internal battery of the phone and the phone's internal battery connectors. The meter's maximum sampling rate is 100 Hz. 

\subsection{Software}
Our software framework includes a data logger, hardware component controller and battery monitor. The data logger samples hardware settings and utilization data such as CPU frequency and load, screen brightness and network traffic. The controller's main job is to create test scenarios. It activates, deactivates and applies workloads on hardware components. For example, while profiling the screen, it changes and fixes its brightness, as well as it turns off other components and fixes the CPU frequency.

The battery monitor records the power consumption data such as the remaining energy, voltage and drawn current during each test session. Accessing the battery chip's values can be done through the battery API, such as Android's BatteryManager class. This API periodically broadcasts these values. 

\section{Experiment}
In this paper we report the results of four experimental scenarios used to validate the battery fuel gauge precision and the accuracy with respect to an external meter. In the first two scenarios, the hardware controller sets the CPU to maximum frequency and the screen brightness to 100\%. The device is held idle for five minutes, then the controller turns off the screen and locks the CPU frequency to prevent the operating system from changing its settings. This second phase (scenario) lasts for five minutes as well. During the experiment, the battery monitor records battery's data (e.g. voltage, current, temperature and remaining energy in the battery) provided by the fuel gauge and the data logger keeps track of system data such as the CPU workload. In addition, a complete system report is collected using the Android Debug Bridge utility. In addition to the screen experiment, we have conducted experiments to collect Wi-Fi and 4G energy consumption data. Three more settings are added as following. After recharging the device and setting up the CPU and screen, the controller downloads a 15~MB file ten times. In between each download job, there is a one-minute rest period. During this period, The controller deactivates the network provider under test for 20 seconds and then reactivates it. As a result, we can capture the energy cost of such events. During each download, the system and battery data as well as network's statistics are recorded, for example, the number of downloaded and uploaded packets and signal strength.

\section{Results}

Before coming to the results, we need to note that we believe the Nexus 6 does not directly supply the remaining energy directly, although the Android documentation claims it. We multiplied the results obtained from the system by the voltage observed for the particular sample taken. The resulting values are then in the correct order of magnitude.

Table~\ref{tab:summary} summarizes the observed energy consumption for our two devices in four different experiments. It appears that the internal meter in the Nexus 9 tends to underestimate the energy consumption while the Nexus 6 fuel gauge overestimates it. 

\renewcommand{\arraystretch}{1.1}
\setlength{\tabcolsep}{3.3mm} 
\begin{table}[t]
	\centering
    \caption{Summary of four experiments on two devices. CPU core speed was set at max frequency. Values are in Wh.}
    \label{tab:summary}
	\begin{tabular}{|l|c|c|c|c|}
		\hline
        \multicolumn{1}{|c|}{}                            & \multicolumn{2}{l|}{Nexus 9} & \multicolumn{2}{l|}{Nexus 6} \\ \hline
		\multicolumn{1}{|c|}{\multirow{2}{*}{Experiment}} & Internal      & External     & Internal      & External     \\ \cline{2-5} 
		\multicolumn{1}{|c|}{}                            & \multicolumn{4}{c|}{Total Energy Measured in Wh}                   \\ \hline
		Screen on                                         & 0.356      & 0.394     & 0.215      & 0.183     \\ \hline
		Screen off                                        & 0.067      & 0.074     & 0.034      & 0.030     \\ \hline
		Wi-Fi download                                    & 1.570      & 1.699     & 0.526      & 0.514     \\ \hline
		4G download                                       & 1.693      & 1.865     & 0.889      & 0.869     \\ \hline
\ignore{
		\multicolumn{1}{|c|}{}                            & \multicolumn{2}{l|}{Nexus 9} & \multicolumn{2}{l|}{Nexus 6} \\ \hline
		\multicolumn{1}{|c|}{\multirow{2}{*}{Experiment}} & Internal      & External     & Internal      & External     \\ \cline{2-5} 
		\multicolumn{1}{|c|}{}                            & \multicolumn{4}{c|}{Total Energy Measured in Wh}                   \\ \hline
		Screen on                                         & 0.356373      & 0.393708     & 0.214601      & 0.183267     \\ \hline
		Screen off                                        & 0.067464      & 0.073694     & 0.033575      & 0.029731     \\ \hline
		Wi-Fi download                                    & 1.570494      & 1.699258     & 0.525549      & 0.513646     \\ \hline
		4G download                                       & 1.692522      & 1.864613     & 0.888606      & 0.869155     \\ \hline
        }
	\end{tabular}
\end{table}

\begin{figure}[h]
	\centering
	\begin{tabular}{@{}c@{}c@{}}
		\includegraphics[clip,trim= 20mm 113mm 20mm 128mm, width=95mm]{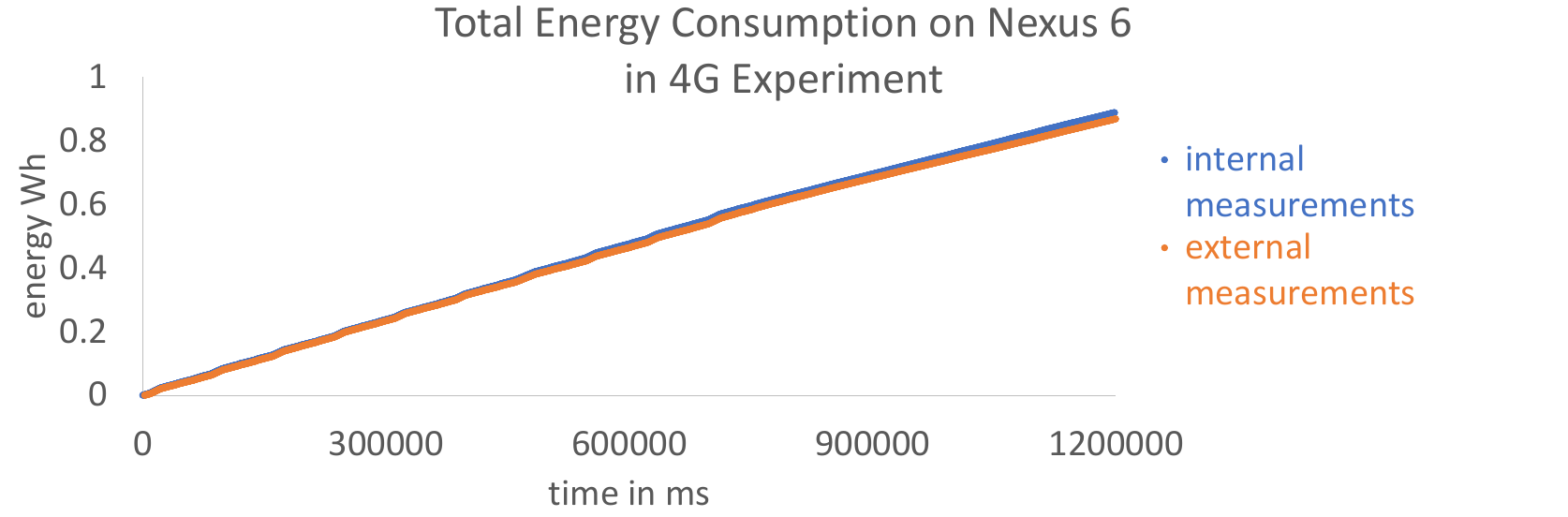}
		%\\
		%\includegraphics[clip,trim= 20mm 112mm 20mm 127mm, width=90mm]{fig/4G-Experiment-on-Nexus-9.pdf}\\
	\end{tabular}\vspace{-2mm}
	\caption{4G experiment on Nexus 6. Energy consumption is measured using internal and external energy meter.}
	\label{fig:totalEnergy}
\end{figure}

\begin{figure}[h]
	\centering
	\begin{tabular}{@{}c@{}c@{}}
		\includegraphics[clip,trim= 20mm 112mm 20mm 127mm, width=90mm]{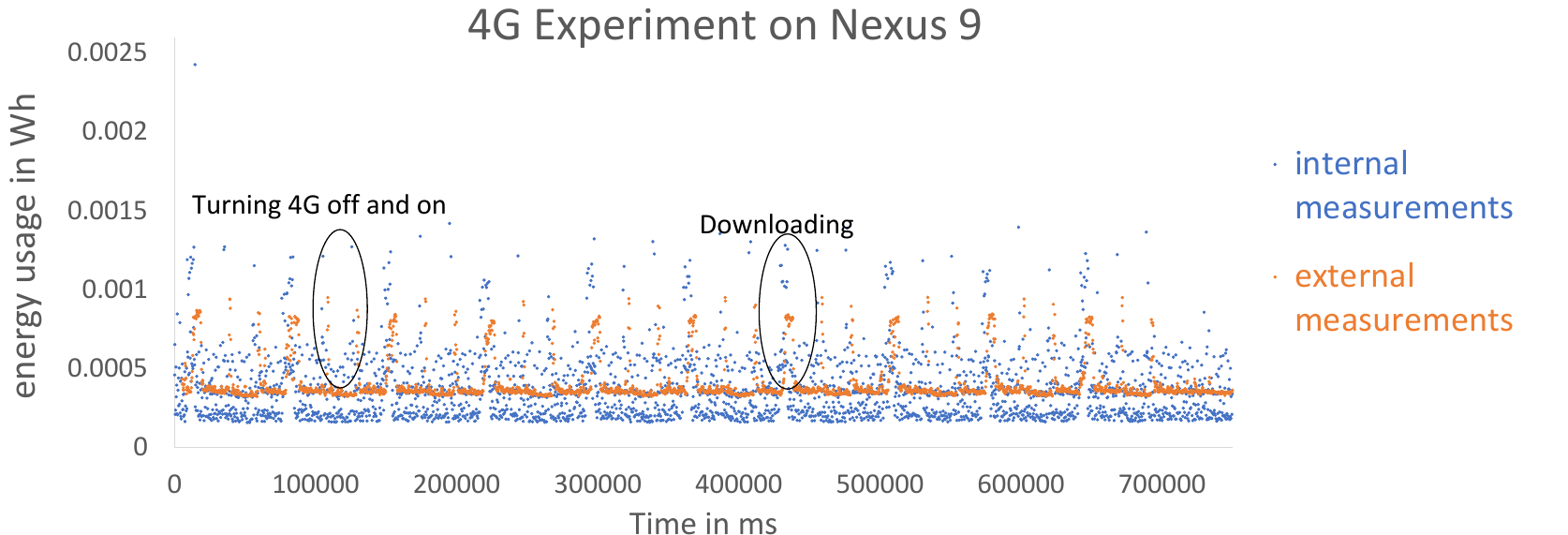}\\
		\\
		\includegraphics[clip,trim= 20mm 112mm 20mm 126mm, width=90mm]{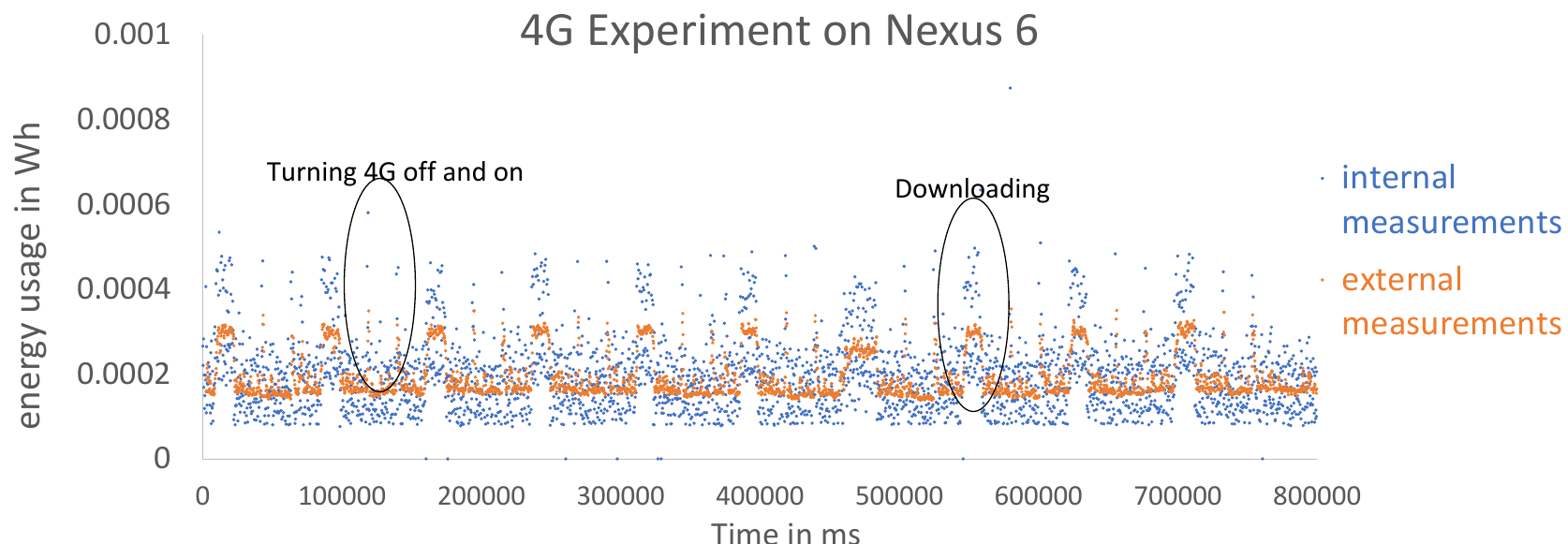}
	\end{tabular}\vspace{-2mm}
	\caption{Wi-Fi experiment on both devices. Energy consumption is measured using internal and external energy meter.}
	\label{fig:4gExperiment}
\end{figure}

Figures \ref{fig:totalEnergy} and \ref{fig:4gExperiment} illustrate different aspects of the experiments. Figure~\ref{fig:totalEnergy} shows the total consumption over time in the 4G experiment and the corresponding individual samples. Both measurements are highly correlated. The external meter is very stable in its readings when the load stays stable. In contrast to this, the internal meter's readings are very noisy during short periods of times (see Figure\ref{fig:4gExperiment}). However, when the considered time window is tens of seconds, then we consider the effective power consumption signal that we get from the internal meter as being stable.  

\ignore{
\begin{figure}[h]
	\centering
	\begin{tabular}{@{}c@{}c@{}}
		\includegraphics[clip,trim= 20mm 108mm 20mm 127mm, width=90mm]{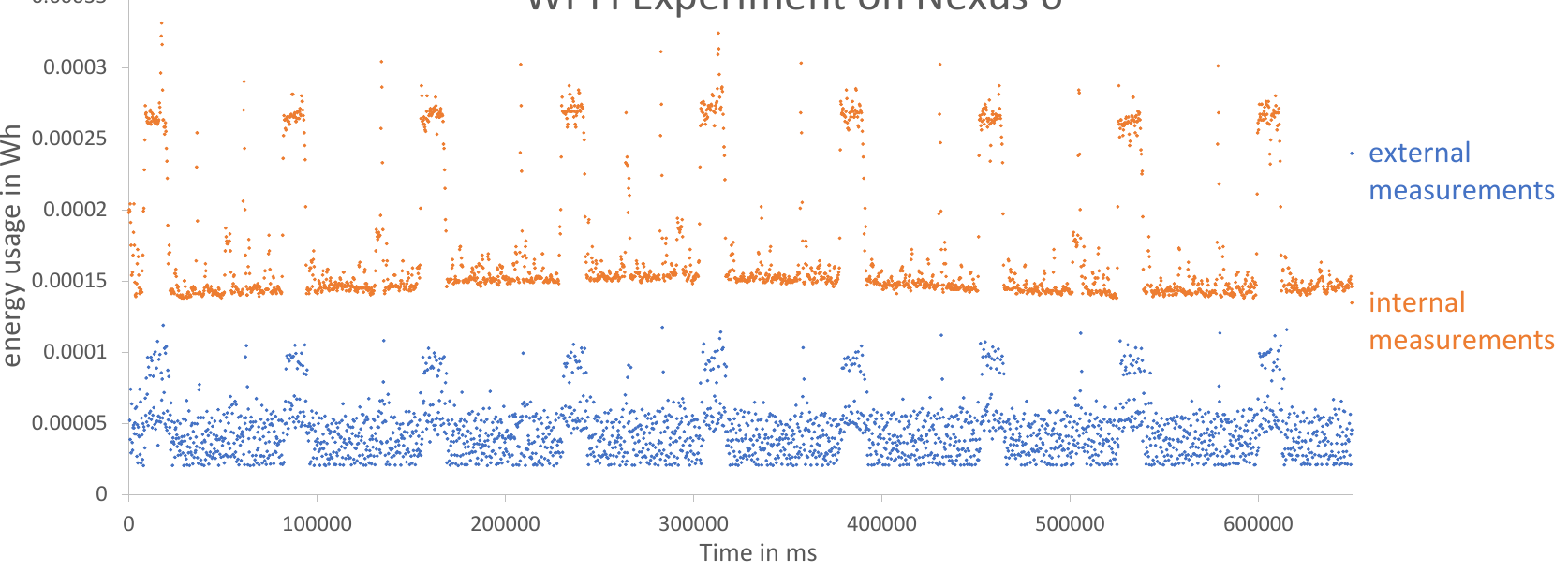}\\
		\\
		\includegraphics[clip,trim= 20mm 116mm 20mm 130mm, width=90mm]{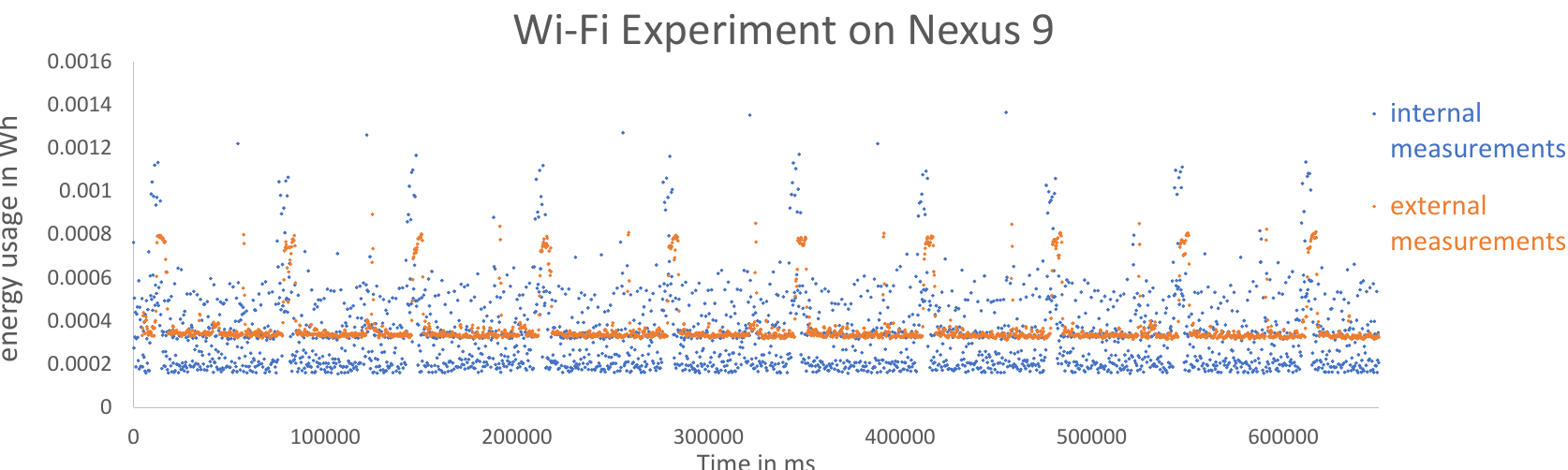}
	\end{tabular}\vspace{-2mm}
	\caption{Wi-Fi experiment on both devices. Energy consumption is measured using internal and external energy meter.}
	\label{fig:Test}
\end{figure}
}

\section{Conclusion and Future Work}

In this article we outlined our approach towards achieving precise power measurements on smart-phones without the use of external meters. 
We require this to (i) create customized power models online and to (ii) optimize the multi-objective energy consumption of software using genetic program improvement~\cite{Bokhari2016}.

Our next step will be to characterize the noise and to investigate the minimal event that we can reliably detect, which will depend on the observation time window and the effect size. In parallel, we will conduct a wide range of experiments on multiple phones, and subsequently create and validate device-specific models using machine-learning techniques, before moving to the fully-automatic optimization of non-functional property optimization directly on smart-phones. 

\bibliographystyle{IEEEtran}
\bibliography{IEEEabrv,library}

\begin{thebibliography}{1}
\providecommand{\url}[1]{#1}
\csname url@samestyle\endcsname
\providecommand{\newblock}{\relax}
\providecommand{\bibinfo}[2]{#2}
\providecommand{\BIBentrySTDinterwordspacing}{\spaceskip=0pt\relax}
\providecommand{\BIBentryALTinterwordstretchfactor}{4}
\providecommand{\BIBentryALTinterwordspacing}{\spaceskip=\fontdimen2\font plus
\BIBentryALTinterwordstretchfactor\fontdimen3\font minus
  \fontdimen4\font\relax}
\providecommand{\BIBforeignlanguage}[2]{{%
\expandafter\ifx\csname l@#1\endcsname\relax
\typeout{** WARNING: IEEEtran.bst: No hyphenation pattern has been}%
\typeout{** loaded for the language `#1'. Using the pattern for}%
\typeout{** the default language instead.}%
\else
\language=\csname l@#1\endcsname
\fi
#2}}
\providecommand{\BIBdecl}{\relax}
\BIBdecl

\bibitem{falaki:diversityUsagePattern}
H.~Falaki, R.~Mahajan, S.~Kandula, D.~Lymberopoulos, R.~Govindan, and
  D.~Estrin, ``Diversity in smartphone usage,'' in \emph{Proceedings of the 8th
  International Conference on Mobile Systems, Applications, and Services}, ser.
  MobiSys '10.\hskip 1em plus 0.5em minus 0.4em\relax ACM, 2010, pp. 179--194.

\bibitem{Shye:utilization1}
A.~Shye, B.~Scholbrock, and G.~Memik, ``Into the wild: Studying real user
  activity patterns to guide power optimizations for mobile architectures,'' in
  \emph{Symposium on Microarchitecture}, ser. MICRO 42.\hskip 1em plus 0.5em
  minus 0.4em\relax ACM, 2009, pp. 168--178.

\bibitem{energyConsumptionBook}
S.~Tarkoma, M.~Siekkinen, E.~Lagerspetz, and Y.~Xiao, \emph{Smartphone energy
  consumption: modeling and optimization}.\hskip 1em plus 0.5em minus
  0.4em\relax Cambridge University Press, 2014.

\bibitem{android:androidManual}
\BIBentryALTinterwordspacing
\emph{Android Power Profiles}, Android, retrieved 03/2016. [Online]. Available:
  \url{https://source.android.com/ devices/tech/power.html}
\BIBentrySTDinterwordspacing

\bibitem{maxim}
\BIBentryALTinterwordspacing
{MAX17047/MAX17050 ModelGauge m3 Fuel Gauge}. Maxim integrated. Viewed November
  2016. [Online]. Available:
  \url{https://datasheets.maximintegrated.com/en/ds/MAX17047-MAX17050.pdf}
\BIBentrySTDinterwordspacing

\bibitem{yoctowatt}
\BIBentryALTinterwordspacing
{Yocto-Watt}. Yoctopuce. Viewed November 2016. [Online]. Available:
  \url{http://www.yoctopuce.com/EN/products/usb-electrical-sensors/yocto-watt}
\BIBentrySTDinterwordspacing

\bibitem{Bokhari2016}
M.~Bokhari and M.~Wagner, ``Optimising energy consumption heuristically on
  android mobile phones,'' in \emph{Proceedings of the 2016 on Genetic and
  Evolutionary Computation Conference Companion}, ser. GECCO '16
  Companion.\hskip 1em plus 0.5em minus 0.4em\relax ACM, 2016, pp. 1139--1140.

\end{thebibliography}
%\begin{thebibliography}{1}

%\end{thebibliography}

\end{document}